\documentclass[conference]{IEEEtran}
\IEEEoverridecommandlockouts
\pdfoutput=1
\usepackage{cite}
\usepackage{amsmath,amssymb,amsfonts}
\usepackage{algorithmic}
\usepackage{graphicx}
\usepackage{tabularx}
\usepackage{textcomp}
\usepackage{xcolor}
 \usepackage{multirow}
\def\BibTeX{{\rm B\kern-.05em{\sc i\kern-.025em b}\kern-.08em
    T\kern-.1667em\lower.7ex\hbox{E}\kern-.125emX}}
\begin{document}

\title{{Traditional Transformation Theory Guided Model for Learned Image Compression}\\
\thanks{This work is supported by the National Natural Science Foundation of
China (No.61627811); Natural Science Foundation of Shaanxi Province(No.2021JZ-04); Joint project of key R \& D universities in Shaanxi Province(No.2021GXLH-Z-093, No.2021QFY01-03).(\emph{Corresponding author: Chenyang Ge.})}
\thanks{The authors are with the Institute of Artificial Intelligence and Robotics, Xi'an Jiaotong University, Xi'an 710049, China (e-mail: lizhiyuan2839@stu.xjtu.edu.cn; cyge@mail.xjtu.edu.cn; 1240231257@qq.co\\m.}}

\author{Zhiyuan Li, Chenyang Ge, Shun Li}

\maketitle

\begin{abstract}
Recently, many deep image compression methods have been proposed and achieved remarkable performance. However, these methods are dedicated to optimizing the compression performance and speed at medium and high bitrates, while research on ultra low bitrates is limited. In this work, we propose a ultra low bitrates enhanced invertible encoding network guided by traditional transformation theory, experiments show that our codec outperforms existing methods in both compression and reconstruction performance. Specifically, we introduce the Block Discrete Cosine Transformation to model the sparsity of features and employ traditional Haar transformation to improve the reconstruction performance of the model without increasing the bitstream cost.

\end{abstract}


\section{Introduction}
With the explosive growth of image data, lossy image compression has become an active research in the field of computer vision for efficient storage and transmission. To this end, traditional image compression methods, such as JPEG \cite{JPEG}, JPEG2000 \cite{JPEG2000}, Webp\cite{WebP} and BPG\cite{BPG}, have been developed and are widely employed in practice. However, these methods often introduce severe blocking artifacts due to their block-based processing, especially under extremely low bitrates.


Recently, with rapid advancements in deep learning, significant advances have been witnessed in neural image compression. Many learning-based image compression algorithms have been proposed\cite{RNN1,VAE1,VAE6}. In contrast to traditional methods which rely on handcrafted rules, these learning-based methods often develop deep neural networks for transforming images into latent features, achieving superior performance and reconstruction quality over traditional image compression algorithms.

Transferring images over extremely limited bandwidth is a practical yet challenging task. Most existing learning-based methods focus on medium and high bitrates and have poor reconstruction quality at extremely low bitrates. Agustsson et al. \cite{GAN} present a Generative Adversarial Network (GAN)-based compression method achieveing dramatic bitrate savings. Their results tend to maintain the high-level semantics but with significant deviated details of original inputs. 

Inspired by traditional transformation theory, we propose a ultra low bitrates enhanced invertible encoding network based on \cite{INC}. The network employs Invertible Neural Networks (INNs) with the strictly invertible property to overcome the information loss problems commonly seen in general image compression autoencoders. Inspired by the fact that the Fourier transformation can almost restore the original signal using only high-energy basic signals like direct current (DC), fundamental waves, and third harmonic waves, we introduce Block Discrete Cosine Transformation (BDCT) to model the sparsity of feature space. During the decoding process, only DC information and low-frequency signals with high energy are used for inverse BDCT, greatly improving the rate performance of the model while ensuring high-quality reconstruction. However, the BDCT may introduce block artifacts. To alleviate this problem, we use Haar transformation to conduct multi-level downsampling of the input image. Due to its reversible and multi-resolution analysis characteristics, Haar transformation helps improve the performance of image restoration. Equipped with the above delicate designs, the proposed method maintains a highly invertible structure based on INNs and Haar transformation, while incorporating the advantages of discrete cosine transformation in concentrating energy in the low-frequency interval. This not only ensures excellent image restoration performance but also greatly improves the rate performance and inference efficiency of the model.

Experimental results show that the proposed method outperforms existing learning-based and traditional methods on the self-built dataset and Kodak \cite{Kodak} dataset, with even greater advantages at lower bitrates. The contributions of this paper are:

\begin{enumerate}
\item  We propose a novel ultra low bitrates enhanced invertible encoding network guided by traditional transformation theory, this is a effective fusion of traditional transformation theory and depth image compression.

\item We apply the Block Discrete Cosine Transformation to model feature sparsity, greatly improveing the compression performance of the model. In addition, We further leverage the Haar transformation to enhance the reconstruction capacity of our network.

\item The proposed method achieves high-quality image reconstruction under low bitrates conditions on both Kodak dataset and self-built dataset.
\end{enumerate}

\begin{figure*}[t]%
\centering
\includegraphics[width=0.9\textwidth]{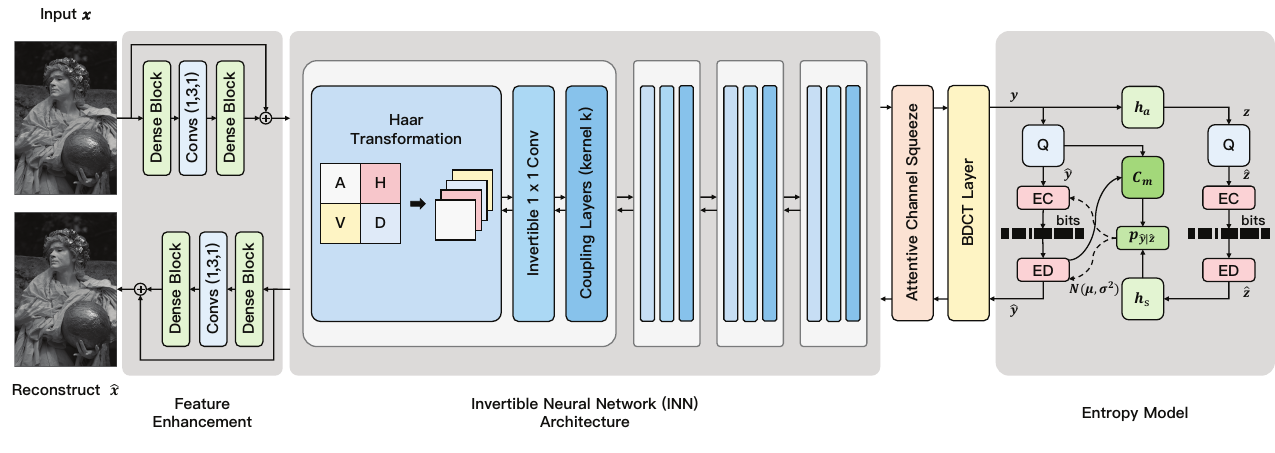}
\caption{Overview of the proposed image compression method}\label{Fig1}
\end{figure*}

\section{Related work}\label{sec2}

\subsection{Traditional Image Compression}\label{subsec21}

Traditional compression standards, such as JPEG \cite{JPEG}, JPEG2000 \cite{JPEG2000}, WebP \cite{WebP} and BPG\cite{BPG} follow a pipeline of transformation, quantization, and entropy coding. Transformation often uses handcrafted modules that incorporate prior knowledge, like Fast Fourier Transform (FFT) and Discrete Cosine Transformation (DCT). Entropy coders include Arithmetic coders, Huffman coder and other coding methods. Some methods like VVC\cite{VVC} introduce intra prediction to improve rate performance. Despite these advances, traditional methods still have limitations due to block compression resulting in blocking artifacts in reconstructed images.

\subsection{Learned Image Compression}\label{subsec22}
In recent years, many learning-based image compression methods have been proposed and achieved impressive performance. Several Recurrent Neural Network (RNN) based methods \cite{RNN1,RNN2,RNN3,RNN4,RNN5} iteratively control image restoration effect, thereby improving rate performance. However, RNN-based methods can demand significant memory and computational resources, making them less suitable for resource-constrained devices or real-time applications. Another field of learning-based image compression methods relies on variational autoencoders (VAEs). Several preliminary works \cite{ETEIC} tackled challenges of bitrate estimation and non-differential quantization, establishing foundations for end-to-end optimization strategies minimizing image distortion and estimated bitrates Simultaneously. Subsequently, numerous VAE-based image compression methods \cite{VAE1,VAE2,VAE3,VAE4,VAE5,ETEIC,VAE6,VAE7,VAE8,VAE9,VAE11} were proposed and achieved impressive performance, but still need improvement under ultra low bitrates. In this article, we introduce traditional transformation theory into learning-based image compression method to achieve better performance under ultra low bitrates.


\section{Methods}\label{sec3}
\subsection{Overview}\label{subsec31}

Building on insights from \cite{INC}, we utilize traditional discrete Haar transformation and BDCT to enhance rate performance of our model while maintaining image reconstruction quality, computational complexity, and parameter efficiency. An overview of our proposed ultra low bitrates enhanced invertible encoding network is illustrated in Fig. \ref{Fig1}. The feature enhancement module bolsters the network's nonlinear representation capacity. The INN architecture mitigates information loss during image compression, improving restoration quality. The attentive channel squeeze layer and the BDCT layer help to remove channel and spatial redundancy, respectively. Finally, the entropy model encodes and decodes features.

The remaining parts are organized as follows: In Section 3.2, we introduce the INN architecture. Then we introduce the BDCT layer to improve the rate performance of the proposed model in Section 3.3. In Section 3.4, we describe the entropy model and its implementation.

\subsection{INN Architecture}\label{INN}
Inspired by Real-NVP \cite{Real-NVP}, we develop an INN architecture using multiple invertible blocks to obtain compact features. Each block comprises a downscaling layer and three coupling layers. The downscaling layer involves a discrete Haar transformation layer and an invertible 1×1 convolution \cite{invc1x1}, reducing input resolution by 2 and quadrupling the channel dimension. We employ the affine coupling layer design from Real-NVP \cite{Real-NVP} for the coupling layer, ensuring strict invertibility of information flow during bidirectional transmission. Following existing methods \cite{Ballé,VAE1,VAE6}, we use 4 invertible blocks in our INN architecture.


\subsubsection{Haar Transformation}\label{Haar transformation}
In each downscaling layer, a Haar transformation layer is applied to split input images into an approximate low-pass representation and three high-frequency residuals \cite{HF1,HF2,HF3}. Specifically, given an input raw image or feature map with height $H$, width $W$, and channel $C$, the Haar transformation layer transforms it into a tensor of shape $(\frac{H}{2} \times \frac{W}{2} \times 4C)$. The first $C$ slices represent the low-pass representation equivalent to Bilinear interpolation downscaling. The other three groups of $C$ slices correspond to high-frequency residuals in the vertical, horizontal and diagonal directions respectively. By separating low- and high-frequency information, the model can preserve sufficient high-frequency components for reconstruction while scaling the image.

\subsubsection{Coupling Layer}\label{Coupling Layer}
For the coupling layer in the INN architecture, given the input $u^{(1:C)}$ with a dimensional size of $C$, it splits the input at position $c(0<c<C)$ into two parts: $u_1(u^{(1:c)})$ and $u_2(u^{(c:C)})$. It performs feature transformations on both two parts, then combines them to yield a $C$ dimensional output $v^{(1:C)}$:
\begin{equation}
\left\{\begin{array}{l}
v_1=u_1 \odot \exp \left(s_2\left(u_2\right)\right)+t_2\left(u_2\right), \\
v_2=u_2 \odot \exp \left(s_1\left(v_1\right)\right)+t_1\left(v_1\right),
\end{array}\right.
\label{eq.1}
\end{equation}
where $\odot$ denotes the Hadamard product, $\exp$ denotes the exponential function, and $s_1$, $s_2$, $t_1$, $t_2$ are arbitrary functions. Symmetrically, taking $v^{(1:C)}$ as input with splitting position $c$, the inverse transformation is easily computed:
\begin{equation}
\left\{\begin{array}{l}
u_2=\left(v_2-t_1\left(v_1\right)\right) \odot \exp \left(-s_1\left(v_1\right)\right), \\
u_1=\left(v_1-t_2\left(u_2\right)\right) \odot \exp \left(-s_2\left(u_2\right)\right),
\end{array}\right.
\label{eq.2}
\end{equation}
Eq. (\ref{eq.1}) and (\ref{eq.2}) show the invertibility is inherently guaranteed by the ingenious mathematical design. Note that the functions $s_1$, $s_2$, $t_1$, and $t_2$ can be arbitrary functions, and need not be invertible. In our implementation, we use the same bottleneck structure as in \cite{INC}.

\subsection{Block Discrete Cosine Transformation}\label{BDCT}
\begin{figure}[t]%
\centering
\includegraphics[width=\linewidth]{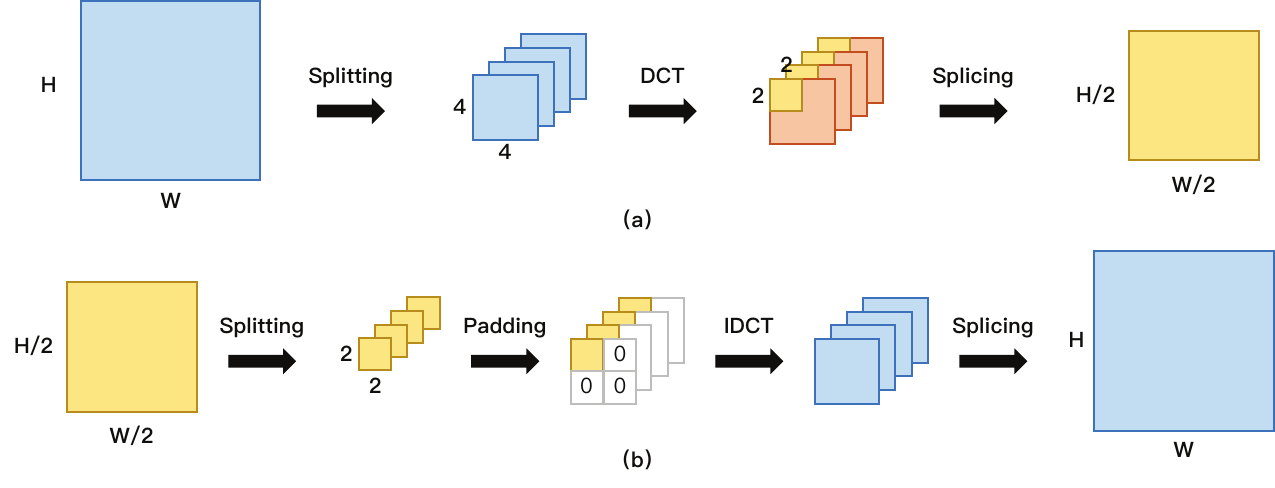}
\caption{Illustration for Block Discrete Cosine Transformation and its inverse process. (a) is the process of BDCT, and (b) is the inverse process of BDCT}\label{Fig2}
\end{figure}

To improve rate performance, we introduce a BDCT layer to model the sparsity of the feature space. As shown in Fig. \ref{Fig2}, given an input tensor of shape $(C, H, W)$, the BDCT layer first divides it into small blocks of size $(C, 4, 4)$. Then it performs discrete cosine transformation operation on these small blocks and the top-left $2\times 2$ portion is utilized for following encoding and decoding operation, i.e. only DC information and low-frequency signals are retained. For the inverse process, the BDCT layer fills the empty positions in small blocks with zeros, and then performs inverse discrete cosine transformation on the filled blocks. Finally, the BDCT layer splices all the blocks into a $(C,H,W)$ tensor. Due to the ignored high-frequency signals contains less information, our method can easily utilize the nonlinear network to compensate for lost information.

\begin{figure*}[t]%
\centering
\includegraphics[width=0.8 \textwidth]{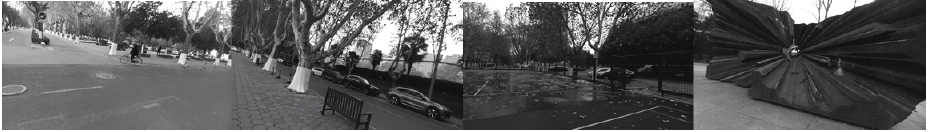}
\caption{Example images in our self-built dataset}\label{Fig3}
\end{figure*}

\begin{figure*}[t]%
\centering
\includegraphics[width=0.72 \textwidth]{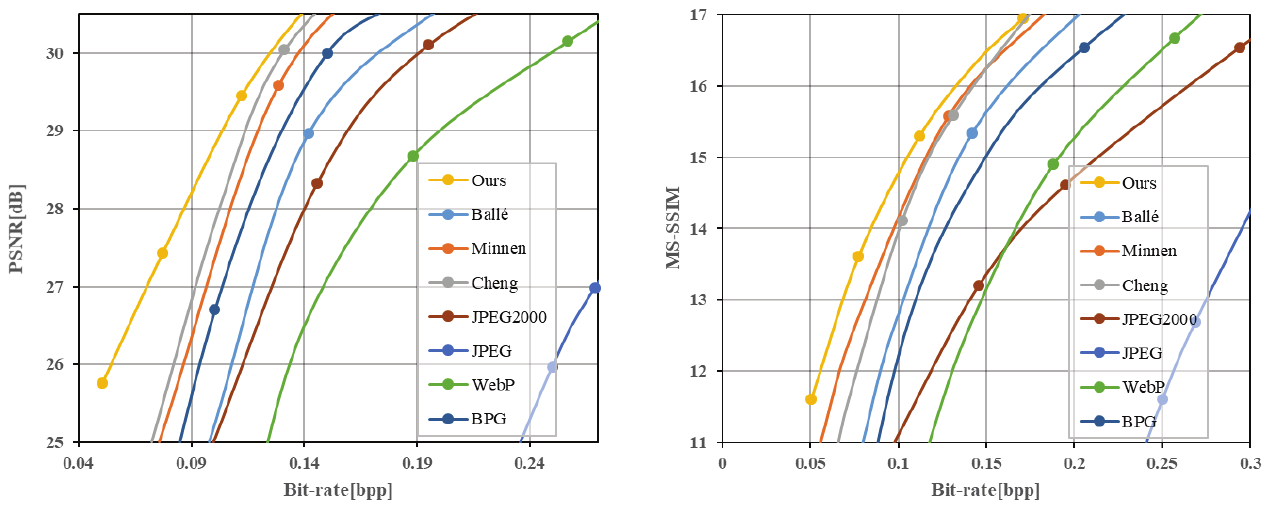}
\caption{Performance evaluation on the self-built dataset}\label{Fig4}
\end{figure*}

\begin{figure*}[t]%
\centering
\includegraphics[width=0.72 \textwidth]{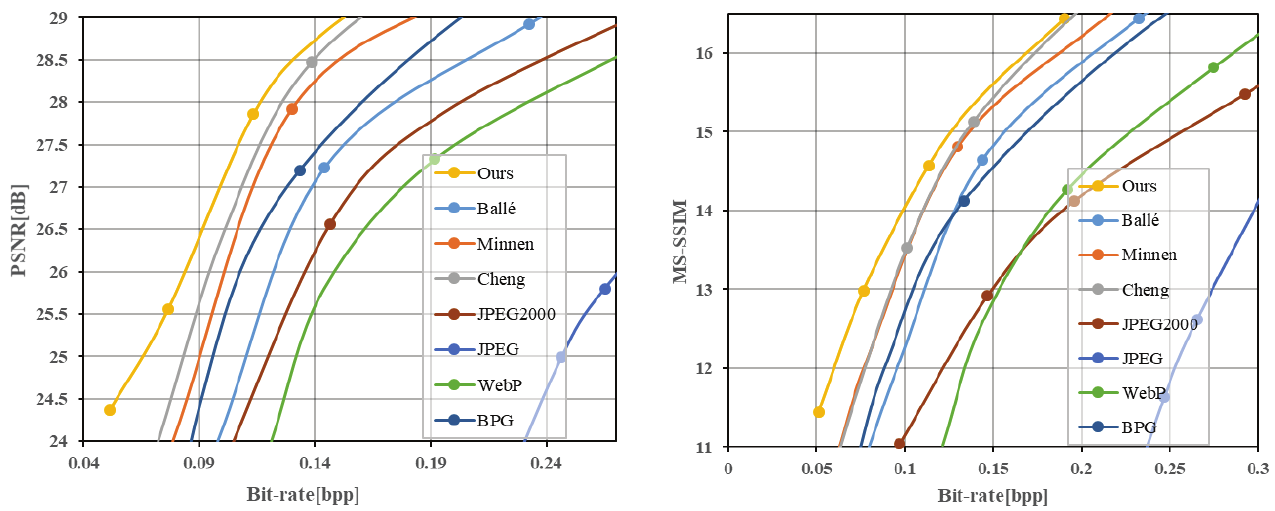}
\caption{Performance evaluation on the Kodak dataset}\label{Fig5}
\end{figure*}


\subsection{Entropy Model}\label{entropy model}
We adopt the hyperprior proposed in \cite{VAE1}, which uses a mean and scale Gaussian distribution to parameterize the quantized latent features $\hat y$, with analysis $h_a$ and synthesis $h_s$ transformations. Since \cite{VAE1} only provides training ideas without inference paradigms, we refer to the inference process of the hyperprior model in \cite{Ballé} and use the asymmetric numeral systems \cite{ANS} to encode data losslessly. 

\section{Experiments}
\subsection{Experimental Setup}
\subsubsection{Datasets}
We evaluate our proposed method on our self-built dataset and the Kodak dataset. To create the self-built dataset, we use an AR0230CS HD camera to collect 10,974 high-quality natural images with a resolution of  $1920\times 1080$. Several example images are shown in Fig. \ref{Fig3}. We randomly select 109 images as the testing set, while the remaining images are used for model training.

\subsubsection{Training details}
To achieve a better RD trade-off, we regularize the training of the proposed method as:
\begin{equation}
    L=\lambda D + R= \lambda d(x,\hat x) + R_{\hat y} + R_{\hat z},
\end{equation}
where $d(x,\hat x)$ represents the distortion between the input image and the reconstructed image, with $d(\cdot)$ referring to mean square error (MSE) or $1 - $MS-SSIM \cite{MS-SSIM}. $\lambda$ is a hyper-parameter controlling the rate-distortion trade-off. $R_{\hat y}$ and $R_{\hat z}$ refer to the entropy of the latent features $\hat y$ and the hyperprior parameters $\hat z$, respectively.  Similar to \cite{INC}, we use the channel number $N$ in the attentive channel squeeze layer and the weight factor $\lambda$ as quality parameters. We train four models optimized with the MSE quality metric, selecting $N$ from the set $\{16, 32, 128, 128\}$ and $\lambda$ from the set $\{24, 24, 16, 32\} \times 10^{-4}$. The value of $\lambda$ refers to CompressAI \cite{CompressAI}.

During training, we utilize the CompressAI-PyTorch library \cite{CompressAI} to train our network. The training epoch is set to 400 with a batch size of 8 and a patch size of 256. We adopt the ADAM optimizer \cite{Adam} with an initial learning rate of $1\times10^{-4}$,  which decreases to $4\times10^{-5}$ at epoch 200 and then to $1.6\times10^{-5}$ at epoch 300.

\subsubsection{Test setting}
Following most existing image compression methods, we adopt the widely-used pixel-wise metrics, PSNR and MS-SSIM, to quantify image distortion, and use bits per pixel (bpp) to evaluate the rate performance. We draw rate-distortion (RD) curves to compare the coding efficiency of different methods based on their rate-distortion performance.

\subsection{Comparisons with State-of-the-art Methods}
\subsubsection{Rate-distortion Performance}
We compare our model with state-of-the-art learned image compression models, including those proposed by Ballé et al. \cite{Ballé}, Minnen et al. \cite{VAE1} and Cheng et al. \cite{VAE6}. We also compare our model with traditional image compression codecs such as JPEG \cite{JPEG}, JPEG2000 \cite{JPEG2000}, WebP \cite{WebP} and BPG \cite{BPG}. All methods use official technical documents for reasoning tests. We evaluate performance using the CompressAI evaluation platform.

Fig. \ref{Fig4} and Fig. \ref{Fig5} show the RD curves comparison on the self-built dataset and Kodak dataset, respectively. Following \cite{VAE6}, we convert MS-SSIM to $-10\log(1 - $ MS-SSIM $)$ for clearer comparison. It is observed that when bpp is below 0.2, i.e. in low bitrates scenarios, our method demonstrates much better performance than comparison approaches. As bpp decreases, advantages of our method become more pronounced. Especially when the bpp is close to 0.05, our method achieves 25.7 dB and 24.3 dB PSNR on the self-built dataset and Kodak dataset, respectively, outperforming other methods.

\begin{figure*}[t]%
\centering
\includegraphics[width=0.83\textwidth]{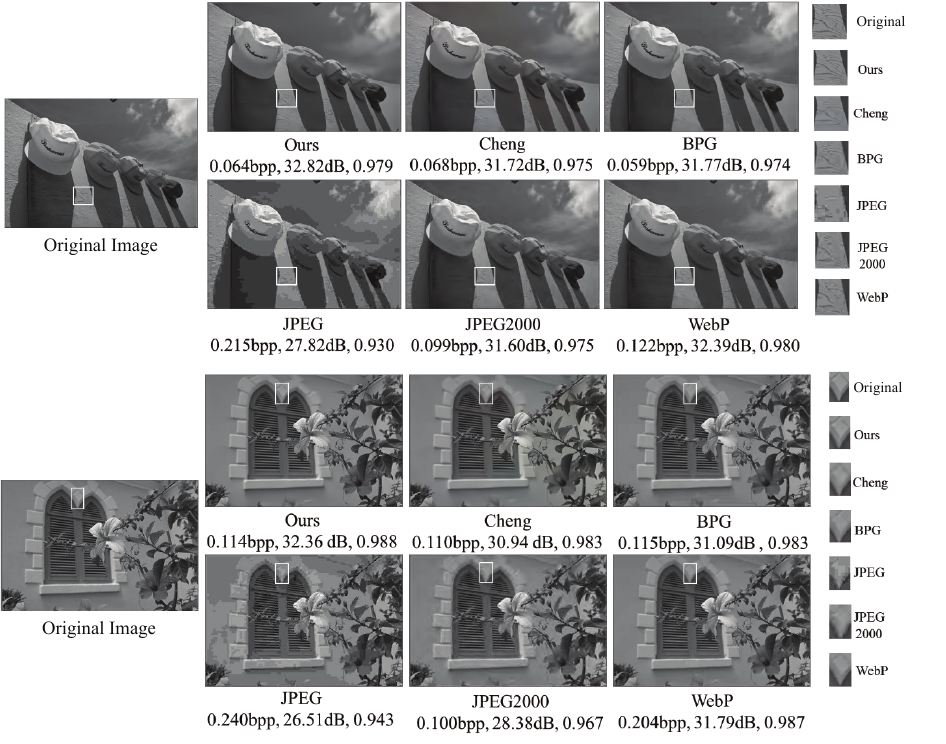}
\caption{Visualization of sample reconstructed images}\label{Fig6}
\end{figure*}

\begin{figure*}[t]
\begin{minipage}[t]{0.49\textwidth}
\centering
\includegraphics[width=\textwidth]{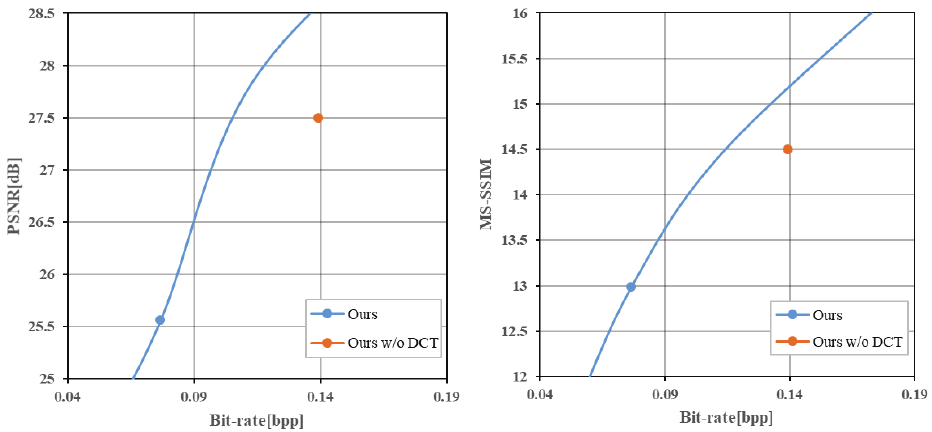}
\caption{Ablation study on BDCT layer.}\label{Fig7}
\end{minipage}
\begin{minipage}[t]{0.49\textwidth}
\centering
\includegraphics[width=\textwidth]{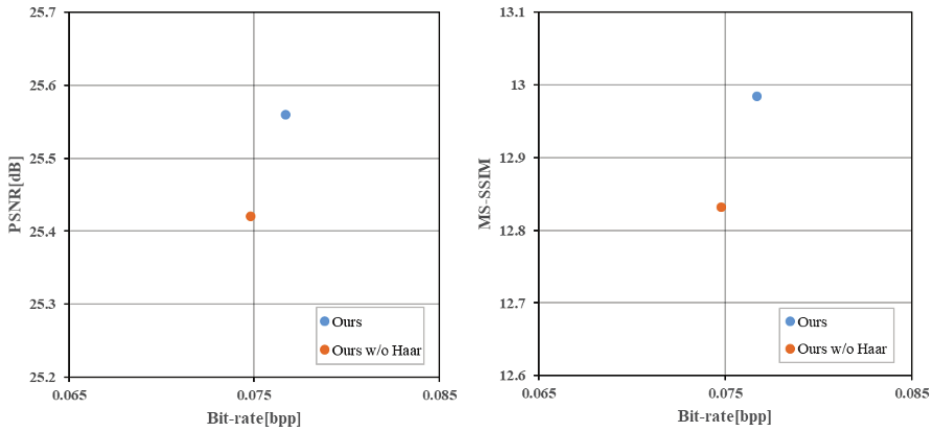}
\caption{Ablation study on Haar transformation layer.}\label{Fig8}
\end{minipage}
\end{figure*}

\subsubsection{Qualitative Results}
Using the model with $N=128$ and $\lambda=0.0016$ as a research subject, we show qualitative comparisons of reconstructed images on the Kodak dataset in Fig. \ref{Fig6}.  We use Cheng et al. \cite{VAE6} as a representative learning-based image compression method and compare our method to traditional methods. As shown in Fig. \ref{Fig6}, our proposed method generates superior results compared to competing approaches. For example, our method is the only one able to effectively restore texture details on the wall, in contrast to other methods that fail to generate plausible details (see first sample). We can also observe our proposed method generates the clearest edge structure in the second sample.

\subsection{Ablation Study}
\subsubsection{BDCT Layer}
We introduce a BDCT layer to model feature space sparsity. To verify its  effectiveness, we conduct an ablation study, removing the BDCT layer and retraining the model with identical settings for fair comparison. Fig. \ref{Fig7} shows rate-distortion points for two models evaluated on Kodak dataset. The BDCT layer significantly reduces required bitrates for storing image content information, with bpp roughly halved. Although image reconstruction quality decreases slightly after adding the BDCT layer, overall compression performance remains superior to model lacking it, i.e. the rate-distortion curve for model with the BDCT layer encloses the rate-distortion point for the model without it.

\subsubsection{Haar Transformation Layer}
To verify the contribution of the Haar transformation layer, we replace it with a pixel shuffling layer and retrain the model with weight factor $\lambda = 0.0024$ and channel number $N = 32$. Fig. \ref{Fig8} shows the rate-distortion points of both models evaluated on the Kodak dataset. We observe that the Haar transformation layer improves image reconstruction quality without increasing bitrate, which shows that Haar transformation effectively promotes the efficient conversion of images from the spatial domain to the feature domain, realizes the more effective data encoding of the image content information, and alleviates the performance degradation caused by the BDCT layer.

\subsubsection{Parameters and Computational complexity}
We further compared the model parameters and computational complexity with respect to the input image size $256 \times 256$, the results are shown in Table \ref{tab1}. It can be seen that adding BDCT layers and Haar transformation layers improve model performance without increasing the model parameters or computational complexity.

\begin{table}[t]
\renewcommand{\arraystretch}{1.1}
\caption{Parameters and Computational complexity with different N}
\begin{center}
\begin{tabular}{c|c|cc}
\hline
\makebox[0.13\linewidth][c]{N} & \makebox[0.13\linewidth][c]{Model} & \makebox[0.2\linewidth][c]{Param(M)} & \makebox[0.15\linewidth][c]{MACs(G)} \\
\hline
\multirow{2}{*}{16} &Baseline$^{\mathrm{a}}$ &4.76 &43.02 \\
                    &Ours &4.76 &43.01 \\
\hline
\multirow{2}{*}{32} &Baseline &4.97 &43.04 \\
                    &Ours &4.97 &43.02 \\
\hline
\multirow{2}{*}{128} &Baseline &9.28 &43.61 \\
                    &Ours &9.28 &43.29 \\
\hline
\multicolumn{4}{l}{$^{\mathrm{a}}$Our method without BDCT and Haar transformation layer.}
\end{tabular}
\label{tab1}
\end{center}
\end{table}

\section{Conclusion}
In this paper, we propose a low bitrates enhanced invertible encoding network guided by traditional transformation theory, which employs the BDCT layer and the Haar transformation layer. The BDCT layer can model the sparsity of features, which is helpful in removing spatial redundancy and improving the rate performance. To improve the reconstruction performance of the proposed method, we apply the Haar transformation layer to equip the model with a certain inductive bias for splitting low- and high-frequency contents. Extensive experiments show our method outperforms state-of-the-art methods in rate-distortion. 

\bibliographystyle{IEEEtran}
\bibliography{main}

\begin{thebibliography}{10}
\providecommand{\url}[1]{#1}
\csname url@samestyle\endcsname
\providecommand{\newblock}{\relax}
\providecommand{\bibinfo}[2]{#2}
\providecommand{\BIBentrySTDinterwordspacing}{\spaceskip=0pt\relax}
\providecommand{\BIBentryALTinterwordstretchfactor}{4}
\providecommand{\BIBentryALTinterwordspacing}{\spaceskip=\fontdimen2\font plus
\BIBentryALTinterwordstretchfactor\fontdimen3\font minus \fontdimen4\font\relax}
\providecommand{\BIBforeignlanguage}[2]{{%
\expandafter\ifx\csname l@#1\endcsname\relax
\typeout{** WARNING: IEEEtran.bst: No hyphenation pattern has been}%
\typeout{** loaded for the language `#1'. Using the pattern for}%
\typeout{** the default language instead.}%
\else
\language=\csname l@#1\endcsname
\fi
#2}}
\providecommand{\BIBdecl}{\relax}
\BIBdecl

\bibitem{JPEG}
G.~K. Wallace, ``The jpeg still picture compression standard,'' \emph{IEEE Transactions on Consumer Electronics}, vol.~38, no.~1, p. xviii–xxxiv, 1992.

\bibitem{JPEG2000}
D.~S. Taubman and M.~W. Marcellin, ``Jpeg2000: Image compression fundamentals, standards and practice,'' \emph{Journal of Electronic Imaging}, vol.~11, no.~2, pp. 286--287, 2002.

\bibitem{WebP}
Google, ``Web picture format,'' 2022.

\bibitem{BPG}
B.~Fabrice, ``{B}{P}{G} image format.'' 2018, \url{https://bellard.org/bpg}.

\bibitem{RNN1}
G.~Toderici, S.~O’Malley, S.~J. Hwang, D.~Vincent, D.~Minnen, S.~Baluja, M.~Covell, and R.~Sukthankar, ``Variable rate image compression with recurrent neural networks,'' \emph{International Conference on Learning Representations}, 2016.

\bibitem{VAE1}
D.~Minnen, J.~Ballé, and G.~Toderici, ``Joint autoregressive and hierarchical priors for learned image compression,'' 2018, in: Proceedings of the Neural Information Processing Systems. pp(10794-10803).

\bibitem{VAE6}
Z.~Cheng, H.~Sun, M.~Takeuchi, and J.~Katto, ``Learned image compression with discretized gaussian mixture likelihoods and attention modules,'' 2020, in: Proceedings of the IEEE/CVF Conference on Computer Vision and Pattern Recognition. pp(7939–7948). \url{https://doi.org/10.1109/cvpr42600.2020.00796}.

\bibitem{GAN}
E.~Agustsson, M.~Tschannen, F.~Mentzer, R.~Timofte, and L.~V. Gool, ``Generative adversarial networks for extreme learned image compression,'' 2019, in: Proceedings of the IEEE/CVF International Conference on Computer Vision. pp(221-231).

\bibitem{INC}
Y.~Xie, K.~L. Cheng, and Q.~Chen, ``Enhanced invertible encoding for learned image compression,'' 2021, {I}n: Proceedings of the 29th ACM International Conference on Multimedia. \url{https://doi.org/10.1145/3474085.3475213}.

\bibitem{Kodak}
E.~K. Company, ``Kodak lossless true color image suite,'' 2013, \url{http://r0k.us/graphics/kodak}.

\bibitem{VVC}
J.~V. E.~T. (JVET), ``Vvc official test model vtm.'' 2021.

\bibitem{RNN2}
G.~Toderici, D.~Vincent, N.~Johnston, S.~J. Hwang, D.~Minnen, J.~Shor, and M.~Covell, ``Full resolution image compression with recurrent neural networks,'' 2017, in: Proceedings of the IEEE conference on Computer Vision and Pattern Recognition. pp(5306-5314). \url{https://arxiv.org/abs/1608.05148}.

\bibitem{RNN3}
N.~Johnston, D.~Vincent, D.~Minnen, M.~Covell, S.~Singh, T.~Chinen, S.~J. Hwang, J.~Shor, and G.~Toderici, ``Improved lossy image compression with priming and spatially adaptive bit rates for recurrent networks,'' 2018, in: Proceedings of the IEEE conference on Computer Vision and Pattern Recognition. pp(4385-4393).

\bibitem{RNN4}
Z.~Wang, A.~Bovik, H.~Sheikh, and E.~Simoncelli, ``Image quality assessment: From error visibility to structural similarity,'' \emph{IEEE Transactions on Image Processing}, vol.~13, no.~4, p. 600–612, 2004.

\bibitem{RNN5}
H.~Zhao, O.~Gallo, I.~Frosio, and J.~Kautz, ``Loss functions for image restoration with neural networks,'' \emph{IEEE Transactions on Computational Imaging}, vol.~3, no.~1, p. 47–57, 2016.

\bibitem{ETEIC}
J.~Ballé, V.~Laparra, and E.~P. Simoncelli, ``End-to-end optimized image compression,'' 2017, in: Proceedings of the International Conference on Learning Representations. pp(1-12).

\bibitem{VAE2}
Z.~Guo, Y.~Wu, R.~Feng, Z.~Zhang, and Z.~Chen, ``3-d context entropy model for improved practical image compression,'' 2020, in: Proceedings of the IEEE/CVF Conference on Computer Vision and Pattern Recognition Workshops. pp(116–117).

\bibitem{VAE3}
D.~Minnen and S.~Singh, ``Channel-wise autoregressive entropy models for learned image compression,'' 2020, in: Proceedings of the IEEE International Conference on Image Processing. pp(3339–3343). \url{https://doi.org/10.1109/icip40778.2020.9190935}.

\bibitem{VAE4}
Y.~Hu, W.~Yang, and J.~Liu, ``Coarse-to-fine hyper-prior modeling for learned image compression,'' 2020, in: Proceedings of the AAAI Conference on Artificial Intelligence. pp(11013–11020). \url{https://doi.org/10.1609/aaai.v34i07.6736}.

\bibitem{VAE5}
J.~Balle, V.~Laparra, and E.~Simoncelli, ``End-to-end optimization of nonlinear transform codes for perceptual quality,'' 2016, in: Proceedings of Picture Coding Symposium. pp(1-5). \url{https://doi.org/10.1109/pcs.2016.7906310}.

\bibitem{VAE7}
T.~Chen, H.~Liu, Z.~Ma, Q.~Shen, X.~Cao, and Y.~Wang, ``End-to-end learnt image compression via non-local attention optimization and improved context modeling,'' \emph{IEEE Transactions on Image Processing}, vol.~30, no.~1, p. 3179–3191, 2021.

\bibitem{VAE8}
Y.~Zhang, K.~Li, K.~Li, B.~Zhong, and Y.~Fu, ``Residual non-local attention networks for image restoration,'' 2019, in: Proceedings of the International Conference on Learning Representations. pp(1–13).

\bibitem{VAE9}
L.~Zhou, Z.~Sun, X.~Wu, J.~Wu, and Y.~Fu, ``End-to-end optimized image compression with attention mechanism,'' 2019, in: Proceedings of the IEEE Conference on Computer Vision and Pattern Recognition Workshops. pp(1–4).

\bibitem{VAE11}
O.~Rippel and L.~Bourdev, ``Real-time adaptive image compression,'' 2017, in: Proceedings of the International Conference on Machine Learning. pp(2922–2930).

\bibitem{Real-NVP}
L.~Dinh, J.~Sohl-Dickstein, and S.~Bengio, ``Density estimation using real nvp,'' 2017, in: Proceedings of the International Conference on Learning Representations. pp(1–10).

\bibitem{invc1x1}
D.~P. Kingma and P.~Dhariwal, ``Glow: Generative flow with invertible 1x1 convolutions. neural information processing systems,'' 2018, in: Proceedings of the Neural Information Processing Systems. pp(1–10).

\bibitem{Ballé}
J.~Ballé, D.~Minnen, S.~Singh, S.~Hwang, and N.~Johnston, ``Variational image compression with a scale hyperprior,'' 2018, in: Proceedings of the International Conference on Learning Representations. pp(1–13).

\bibitem{HF1}
P.~I. Wilson and J.~Fernandez, ``Facial feature detection using haar classifiers,'' \emph{Journal of Computing Sciences in Colleges}, vol.~21, no.~4, p. 127–133, 2006.

\bibitem{HF2}
R.~Lienhart and J.~Maydt, ``An extended set of haar-like features for rapid object detection,'' 2002, proceedings of the International Conference on Image Processing. pp(900–903).

\bibitem{HF3}
L.~Ardizzone, C.~Lüth, J.~Kruse, C.~Rother, and U.~Köthe, ``Guided image generation with conditional invertible neural networks,'' 2019, \url{https://arxiv.org/abs/1907.02392}.

\bibitem{ANS}
J.~Duda, ``Asymmetric numeral systems,'' 2009, \url{https://arxiv.org/abs/0902.0271}.

\bibitem{MS-SSIM}
Z.~Wang, E.~P. Simoncelli, and A.~C. Bovik, ``Multiscale structural similarity for image quality assessment,'' 2003, in: Proceedings of the Asilomar Conference on Signals, Systems and Computers. pp(1398-1402).

\bibitem{CompressAI}
J.~Bégaint, F.~Racapé, S.~Feltman, and A.~Pushparaja, ``Compressai: a pytorch library and evaluation platform for end-to-end compression research,'' 2020, \url{https://arxiv.org/abs/2011.03029}.

\bibitem{Adam}
D.~P. Kingma and J.~Ba, ``Adam: A method for stochastic optimization,'' 2015, in: Proceedings of the International Conference on Learning Representations. pp(1-11).

\end{thebibliography}

\end{document}